\documentclass[aps,pra,amssymb,twocolumn,showpacs,supersriptaddress,groupeaddress]{revtex4-1}
\usepackage{amssymb}
\usepackage{amsmath}
\usepackage{graphicx}
\usepackage{dcolumn}
\usepackage{bm}
\usepackage{bm}
\usepackage{natbib,hyperref}
\usepackage{hyperref}
\usepackage{color}
\hypersetup{colorlinks=true, 
    linkcolor=red,          
    citecolor=magenta,        
    filecolor=gree,      
    urlcolor=blue           
}

\begin{document}
\title{Time-connected phase slips  in current-driven 
	two-band superconducting wires}

\author{Daniel Dom\'{\i}nguez}

\affiliation{Centro At{\'{o}}mico Bariloche and Instituto Balseiro, 8400 San Carlos de Bariloche, Argentina.\\
}

\author{Jorger Berger}
\affiliation{Department of Physics, Braude College, 2161002 Karmiel, Israel
}



\date{\today}%
\begin{abstract}
	
We study  quasi-one dimensional wires of two-band superconductors driven by an electrical current. 
We find that the onset of dissipation can occur with the nucleation of
 time-connected phase slips (t-PS). 
 The topological structure of the t-PS consists of
two phase slips (one in each order parameter) separated in time and
connected via an interband vortex string along the time direction.
This shows as a two-peak structure in voltage vs. time.
We discuss the conditions for observing t-PS, depending on the interband coupling strength and the relaxation time scales for  each order parameter.

\end{abstract}
\maketitle

\section{Introduction}

Abrikosov vortices \cite{abrikosov} in superconductors are topological singularities where the superconducting order parameter $\Psi=|\Psi|e^{i\theta}$ vanishes at a point in space (actually, in a line of points in three dimensions)  and the phase $\theta$ changes by $2\pi n$ around a closed loop that encircles the vortex, $\oint d\theta = 2\pi n$. The integer $n$ defines the ``vorticity'' associated to the topological singularity. Vortices nucleate in type II superconductors \cite{tinkham} in the presence of a magnetic field $\mathbf{B}$,   which determines the vortex density.
 Another interesting case of topological singularities occurs in quasi one-dimensional superconducting wires. When driven by a current, the onset of voltage usually occurs through the nucleation of phase-slips (PS) \cite{skocpol1974,kramer1977,kramer1978,ivlev1978,ivlev1980,watts-tobin1981,ivlev1984,kopnin2001}.
The superconducting order parameter $\Psi=|\Psi|e^{i\theta}$ vanishes in a point in the wire (the PS center) at a given time, after which it recovers a finite value, and this process repeats periodically. Throughout each of these events in time the phase $\theta$ changes by $2\pi$.  Ivlev and Kopnin \cite{ivlev1978} have shown that the PS are vortices in space-time. Phase increments $d\theta$, integrated along a  closed loop in space and time around a PS give $\oint_{\rm space-time} d\theta = 2\pi$. Thus, PS are topological singularities in two-dimensional space-time. The electric field $E$ (instead of $B$) determines their density in time \cite{ivlev1978}, {\it i.e.} the number of PS per unit time.
The  dynamics of PS and the current-voltage curves   in superconducting wires have been studied in detail both theoretically and experimentally since the  1970s \cite{skocpol1974,kramer1977,kramer1978,ivlev1978,ivlev1980,watts-tobin1981,ivlev1984,kopnin2001,michotte2004}. The  dynamics of PS can be modeled with the time dependent Ginzburg-Landau (TDGL) equation \cite{kramer1977}. There is nowadays a good understanding of the dependence of the PS dynamics on TDGL parameters, wire size and boundary conditions \cite{michotte2004,vodolazov2011,ludac2008,kim2010,baranov2011,baranov2013,kallush2014,berger2015,kimmel2017}. Recent experiments in nanowires have achieved detailed control and manipulation of the PS, with pulses of electrical current \cite{buh2015}, with infrared laser pulses \cite{madan2018}, and with microwave radiation \cite{kato2020}.

In the last two decades, the discovery of multiband superconductivity in MgB$_2$ \cite{choi2002}, 
and later on in materials such as NbSe$_2$\cite{boaknin2003}, OsB$_2$\cite{singh2010},  FeSe$_{0.94}$ \cite{khasanov2010}, LiFeAs \cite{Kim2011}  and  other iron-pnictide superconductors, has stimulated the study of the physics of vortices with nontrivial topological properties \cite{babaev2002}. For instance, two-band superconductors are described with two order parameters $\Psi_j=|\Psi_j|e^{i\theta_j}$, $j=1,2$. Therefore, the topological singularities of the two order parameters have to be considered, $\oint d\theta_1 = 2\pi n_1$ and $\oint d\theta_2 = 2\pi n_2$ \cite{babaev2002,smorgrav2005}. The simplest case consists of composite vortices centered at the same point with $n_1=n_2$ , which are encountered in the equilibrium sates in two-band bulk superconductors under a magnetic field.  The so-called non-composite vortices are topological singularities displaced in space ($n_1\not=n_2$ in each point) which are associated with fractional magnetic flux quanta \cite{babaev2002}. They can appear in finite mesoscopic samples \cite{chibotaru2007,geurts2010,pina2012,silva2014}, at the sample boundaries 
\cite{silaev2011}, and in samples driven at high currents \cite{lin2013,mosquera2017}. 

 In one-dimensional two-band superconducting wires the existence of phase textures and topological solitons of the interband phase difference $\theta_2-\theta_1$ has been extensively studied \cite{tanaka2001,gurevich2003b,gurevich2006,yerin2007,lin2012,fenchenko2012,lin2014,tanaka2015,tanaka2015b,marychev2018,berger2011}. 
In the soliton states the phase difference $\theta_2-\theta_1$ varies with the space coordinate and rotates in multiples of  $2\pi$ along the wire length \cite{tanaka2001}. 
The phase solitons are induced by a finite driving current {below the critical current of the wire}. One nucleation mechanism can be charge imbalance at the boundary between the superconducting wire and a normal lead \cite{gurevich2003b}.  The number of phase solitons depends on the driving current and the wire length \cite{marychev2018}. 

In this work we address the dynamics of two-band superconducting wires at the onset of dissipation, above their critical current. We analyze the topological nature of the induced phase slips and their dependence on interband coupling strength by solving numerically a time dependent two-band Ginzburg-Landau equation. We find that the induced PS in the two bands nucleate at the same place but are {\it separated in time}. They are connected through a topological singularity in the interband phase difference that is oriented along the time direction, and we name these nontrivial topological objects as ``time-connected phase slips" (t-PS).
The paper is organized as follows. In Section II we introduce the time-dependent Ginzburg-Landau equations to be solved, boundary conditions and parameters to be considered. In Section III we report our results for the current-voltage curves and characterize the space and time dependence of the induced PS.
In Section IV the topological nature of the t-PS is described in detail and in Section  V we report their dependence on interband coupling and relaxation parameters. Finally, in Section VI we summarize and discuss our findings.

\section{Model and definitions}

%
%
%
%

For a two-component superconductor, the free energy is ${\cal F}=\int d^3{\bf r}f({\bf r})$ \cite{zhitomirsky2004,gurevich2003,koshelev2005} with

$$
\begin{aligned}
f= &\alpha_{1}\left|\Psi_{1}\right|^{2}+\frac{\beta_{1}}{2}\left|\Psi_{1}\right|^{4}+\frac{1}{2 m_{1}}\left|\left(-i \hbar \nabla-\frac{2 e}{c} \mathbf{A}\right) \Psi_{1}\right|^{2}+\\
&\alpha_{2}\left|\Psi_{2}\right|^{2}+\frac{\beta_{2}}{2}\left|\Psi_{2}\right|^{4}+\frac{1}{2 m_{2}}\left|\left(-i \hbar \nabla-\frac{2 e}{c} \mathbf {A}\right) \Psi_{2}\right|^{2}+\\
&\frac{|\mathbf {B}|^{2}}{8 \pi}-\gamma\left(\Psi_{1} \Psi_{2}^{*}+\Psi_{2} \Psi_{1}^{*}\right)
\end{aligned}
$$
where $\alpha_1,\alpha_2,\beta_1,\beta_2,m_1,m_2$ are the Ginzburg-Landau expansion coefficients,   ${\bf A}$ is the  vector potential, ${\bf B}$ is the magnetic field, and  $\gamma$ is the interband Josephson coupling parameter.

We study  the dynamics near the critical temperature $T\lesssim T_c$ using the time-dependent Ginzburg-Landau equations generalized to a two-band superconductor \cite{vargunin2020,gurevich2003b,gurevich2006,berger2011,fenchenko2012,marychev2018,mosquera2017}:

\begin{eqnarray}        
&&\Gamma_1\left(\frac{\partial}{\partial t}+i\frac{2e}{\hbar} \phi\right) \Psi_{1}=-\frac{\delta{\cal F}}{\delta \Psi_1^*}\nonumber\\	
&&\Gamma_2\left(\frac{\partial}{\partial t}+i\frac{2e}{\hbar} \phi\right) \Psi_{2}=-\frac{\delta{\cal F}}{\delta \Psi_2^*}\\
&&{\bf J}=-\sigma\left(\nabla\phi +\frac{1}{c}\frac{\partial\mathbf{A}}{\partial t}\right) \nonumber\\
&&~~+\frac{2e\hbar}{m_1}\operatorname{Im}\left(\Psi_{1}^{*} [\nabla-i(2e/\hbar c)\mathbf{A}] \Psi_{1}\right) \nonumber\\
&&~~+\frac{2e\hbar}{m_{2}} \operatorname{Im}\left(\Psi_{2}^{*} [\nabla-i(2e/\hbar c)\mathbf{A}] \Psi_{2}\right)\nonumber
\end{eqnarray}
where $\Gamma_i=\frac{\hbar^{2}}{2 m_i D_{i}}$ with $D_{i}$  the diffusion coefficients, $\sigma$ is the normal conductivity, $\phi$ is the electric potential, and ${\bf J}$ the electric current density. We will take ${\bf A}=0$, since  there is no applied external magnetic fields and the effect of the  self-induced magnetic field in the  one-dimensional wire is negligible.

We normalize $\Psi_{i}$ by $\Psi_{10}=\sqrt{-\alpha_{1}/\beta_{1}}$,
length scale by $\xi_{1}=\sqrt{\hbar^{2}/2m_{1}(-\alpha_{1})}$, time
by $t_{0}=4\pi\sigma\lambda_{1}^{2}/c^{2}=\sigma m_{1}\beta_{1}/2e^{2}(-\alpha_{1})$,
electric potential $\phi$ by $v_0=\hbar/2et_{0}$ and current density $J$ by
$J_{0}=\hbar\sigma/2et_{0}\xi_{1}$. With this normalization the TDGL
equations in a one-dimensional wire are:

\begin{eqnarray}
&&\eta_1\left(\frac{\partial}{\partial t}+i \phi\right) \Psi_{1}=\frac{\partial^{2}\Psi_{1}}{\partial x^2} +\left(1-\left|\Psi_{1}\right|^{2}\right) \Psi_{1}+g \Psi_{2}
\label{eqPsi}\\
&&\eta_2\left(\frac{\partial}{\partial t}+i \phi\right) \Psi_{2}=k\frac{\partial^{2}\Psi_{2}}{\partial x^2} +\left(a-b\left|\Psi_{2}\right|^{2}\right) \Psi_{2}+g \Psi_{1}\nonumber \\
\end{eqnarray}
and
\begin{equation}
J=-\frac{\partial\phi}{\partial x} +\operatorname{Im}\left(\Psi_{1}^{*} \frac{\partial\Psi_{1}}{\partial x} \right)+k \operatorname{Im}\left(\Psi_{2}^{*} \frac{\partial\Psi_{2}}{\partial x}\right)
\label{eqJ}
\end{equation}
with $a=\alpha_2/\alpha_1$, $b=\beta_2/\beta_1$, $g=\gamma/|\alpha_1|$, $k=m_1/m_2$, $\eta_i=\Gamma_i/t_0|\alpha_1|$; we also
define $d=D_{1}/D_{2},$ and thus $\eta_{2}=kd\eta_{1}$.

We use the convention that $\Psi_1$ corresponds to the `strong' band
and $\Psi_2$ to the `weak' band and consider the case that both bands are superconducting, meaning that $\alpha_1(T)<\alpha_2(T)<0$, and thus $0<a < 1$. 
Also, for consistency with the `weak' band  choice, we consider parameters such that the second band has smaller equilibrium gap, $(\Psi_{20}/\Psi_{10})^2=a/b<1$ and larger coherence length,
$\xi_2^2/\xi_1^2= k/a>1$. Furthermore, weak interband coupling means $\gamma<|\alpha_2|<|\alpha_1|$, i.e. $g<a<1$.
In contrast to single-band superconductors for which the relaxation constant is $\eta\approx 5.79$ \cite{kramer1977,kramer1978,ivlev1984}, in multiband superconductors the relaxation constants $\eta_{1},\eta_{2}$ depend strongly on system parameters \cite{vargunin2020}. The ratio $\eta_2/\eta_1=kd$ depends on the ratio of diffusivities
$d=D_1/D_2$, which has been found to vary significantly  on different superconducting compounds  \cite{dai2011,vargunin2020} (typically in the range $0.1 \lesssim d \lesssim 10$).
The relaxation times of the order parameters are
$t_i=\Gamma_i/|\alpha_i|$, which in terms of the normalization $t_0$ are $t_1/t_0=\eta_1$ and $t_2/t_0=\eta_2/a$, and the ratio of these time scales is $t_2/t_1=kd/a$.
Here we will focus on the case $\eta_2/a>1$ and $\eta_1<1$, i.e. the weak band order parameter has a large relaxation time while the strong band has fast relaxation.

We integrate numerically Eq.(\ref{eqPsi}) with a semi-implicit Crank-Nicholson algorithm (see Appendix for details), with $\delta t=0.02$ and
$\delta x=0.25$.
We consider superconducting wires of length $L$ with a superconducting bank boundary condition: 
\begin{eqnarray}
\Psi_j(x=0,t)&=&\Delta_je^{-i\int_0^t\phi(0,t')dt'}\label{BC}\\
\Psi_j(x=L,t)&=& \Delta_j e^{-i\int_0^t\phi(L,t')dt'}\nonumber
\end{eqnarray}
where $\Delta_j=\Psi_j^{\rm eq}$ is the equilibrium order parameter of the $j$-band at zero current, obtained numerically by solving the equilibrium   homogeneous equations,
\begin{eqnarray}
	\left(1-|\Delta_1|^{2}\right) \Delta_{1}+g \Delta_{2}&=&0
	\nonumber\\
\left(a-b|\Delta_2|^{2}\right) \Delta_{2}+g \Delta_{1}&=&0\nonumber
\end{eqnarray}
For fixed current density $J$, $\phi(x,t)$ is found by direct integration of  Eq.~(\ref{eqJ}), with the additional knowledge that  $\partial\phi/\partial x =0$ at $x=0,L$. The time dependent voltage drop per unit length is determined as $v(t)=[\phi(0,t)-\phi(L,t)]/L$.

\section{Current-voltage curves and nucleation of phase slips}

\begin{figure}[tbh]
	\includegraphics[width=\columnwidth]{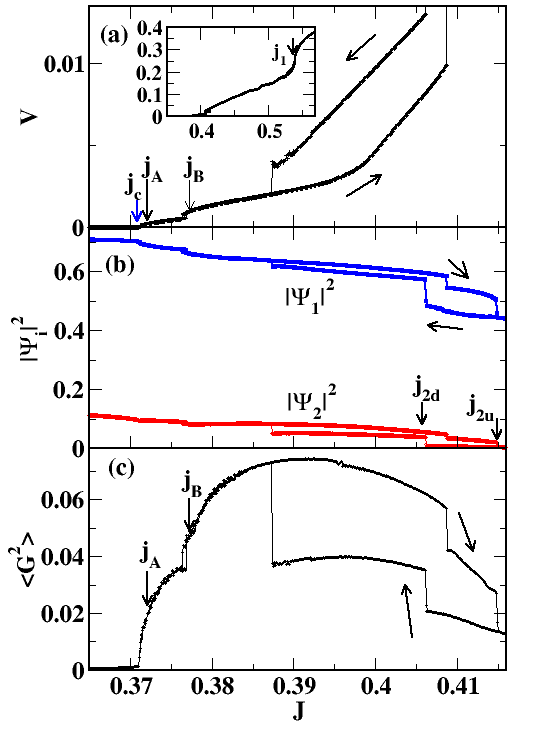}
	\caption{(a) IV curve for a wire of length $L=50\xi_1$.  $V$ is the average voltage per unit length normalized by $v_0/\xi_1$ and the   current $J$ is normalized by $J_0$. The onset of dissipation at $j_c$ is indicated. The inset shows the curve in a wider range of currents, where $j_1$ indicates the current for which the wire becomes normal.
		(b) Average order parameters $|\Psi_1|^2$ (blue line) and 
		$|\Psi_2|^2$ (red line) as functions of $J$.
		(c) Measure of interband phase texture $\langle G^2\rangle$ (see text for definition) as a function of $J$. System parameters: $a=0.2$, $b=1.2$, $k=5.2$, $g=0.08$,
		$\eta_{1}=0.5$, $\eta_{2}=5.2$. The current $j_A$ corresponds to the case shown in Fig.\ref{fig:psi} and $j_B$ corresponds to the case shown in Fig.\ref{fig:psis}(a). In (b) and (c), $|\Psi_1|^2$,  $|\Psi_2|^2$ and $\langle G^2\rangle$ are normalized by $\Psi_{10}^2$.
 		 }
	\label{fig:iv}
\end{figure}

We first calculate the current-voltage (IV) curves for 
$a=0.2$, $b=1.2$, $k=5.2$, $g=0.08$,
$\eta_{1}=0.5$ and $\eta_{2}=5.2$. 
We simulate the dynamics ramping up and ramping down the external current $J$ for wires of length $L=50\xi_1$. We start with the equilibrium values of $\Psi_i(x)$ at $J=0$ and increase gradually the current, taking as initial condition the final values of $\Psi_i(x)$ at the previous current step. After reaching the normal state, we lower the current back to $J=0$, following the same procedure.

\begin{figure}[!ht]
	\includegraphics[width=6.5cm]{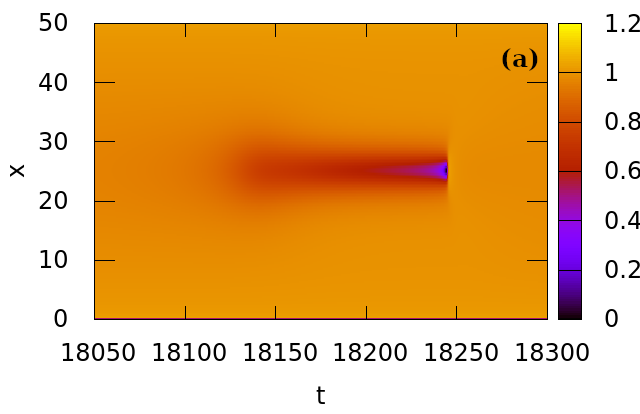}
	\includegraphics[width=6.5cm]{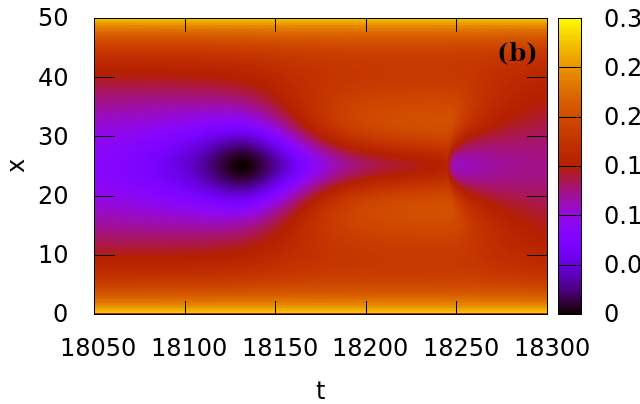}
	\includegraphics[width=6.5cm]{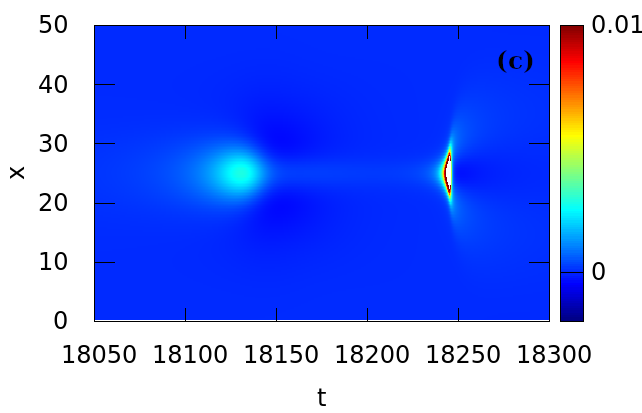}
	\includegraphics[width=6.5cm]{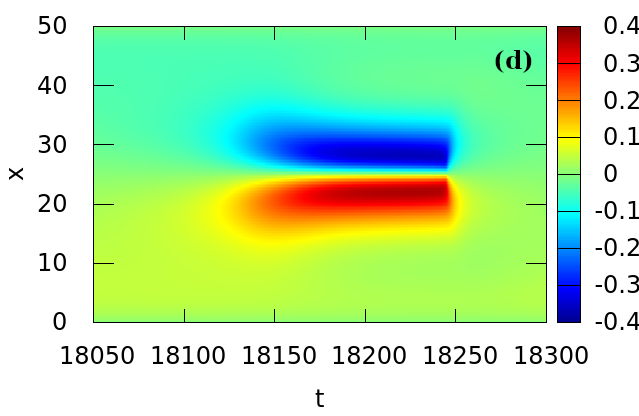}
	\caption{Intensity plots  in space-time coordinates showing the structure of phase slips for a current  $J=0.372\gtrsim j_c$ [indicated as $j_A$ in Fig.\ref{fig:iv}(a)]. Order parameters (a) $|\Psi_1(x,t)|^2$ and (b) $|\Psi_2(x,t)|^2$, normalized by $\Psi_{10}^2$. (c) Electric field  $E(x,t)=-\frac{\partial }{\partial x}\phi(x,t)$ normalized by $v_0/\xi_1$. (d) Interband phase texture $G(x,t)=|\Psi_1||\Psi_2|\sin(\theta_1-\theta_2)$, normalized by $\Psi_{10}^2$. Coordinate $x$ is normalized by $\xi_1$ and time by $t_0$.}
	\label{fig:psi}
\end{figure}

We have calculated numerically
the time-averaged voltage per unit length
$V=\langle v(t)\rangle$ [shown in Fig.\ref{fig:iv}(a)],
and the time and length averaged order parameters
$\langle|\Psi_i|^2\rangle=
\langle\frac{1}{L}\int_0^L  |\Psi_i(x,t)|^2dx\rangle $, $i=1,2$ 
[shown in Fig.\ref{fig:iv}(b)]. 
We find that there is an onset of dissipation at the current density  $J=j_c$.  Above $j_c$ the voltage increases, starting with a steep square root dependence ($V\propto \sqrt{J-j_c}$) \cite{michotte2004,baranov2011} followed by a quasilinear behavior. As we will show, this voltage onset corresponds to the nucleation of one phase slip in each band.
Upon increasing the current there are further onsets of subsequent  regions of quasilinear dependence, that correspond to increasing number of PS centers at more than one position. 
At a larger current, $J=j_{2u}$, the $2$-band becomes normal   while the $1$-band is still superconducting. 
At an even higher current $j_{1} >j_{2u}$ the wire becomes completely normal in the bulk [note that $\Psi_i$ is always finite near the edges due to Eq.(\ref{BC})]. The  current $j_{1}$ is shown in the inset of Fig.\ref{fig:iv}.
As the current decreases, there is hysteresis and the $2$-band stays normal down to lower currents, becoming superconducting at a current $j_{2d}<j_{2u}$. Below $j_{2d}$ the voltage decreases with current, with a different sequence of quasilinear regions, showing a noticeable hysteresis.

We focus in the range $j_c<J<j_{2d}$,  where there is a finite voltage  while both bands are superconducting.
In order to understand the nucleation of PS near the onset of dissipation, $J\gtrsim j_c$ ,  we study the order parameters $|\Psi_1(x,t)|^2$ and $|\Psi_2(x,t)|^2$ as functions of $x$ and $t$.
 We find that PS  nucleate at the center of the wire, where each order  parameter vanishes periodically in time.
 In Fig.\ref{fig:psi}(a) and (b) we show  $|\Psi_1(x,t)|^2$ and $|\Psi_2(x,t)|^2$ in a time window where both PS can be observed. The PS  of the weak band (corresponding to $\Psi_2$) occurs  earlier than the PS  in which $\Psi_1$ vanishes.
The PS vortex core for the $i$-band (region where the order parameter drops) has a length scale of the order of $\xi_i$ and a time scale of the order of $t_i$. Therefore, the core of the $\Psi_2$ phase slip is larger than the core of $\Psi_1$.  
The electric field $E(x,t)=-\partial\phi(x,t)/\partial x$ is plotted in  Fig.\ref{fig:psi}(c). Away from the PS the electric field is zero, $E\approx0$, as expected. There are two maxima of significant $E$ at the cores of the PSs  of each band. The maximum of $E$ in the 2-band is much smaller than the maximum in the 1-band. The characteristic time lag between these two PSs will be denoted  $t_\gamma$ from now on.

Since superconductivity vanishes at different times in the two bands, we expect that the phases $\theta_i$ of the two order parameters will be different. To quantify the occurrence of interband phase textures we evaluate the interband ``Josephson current'':
\begin{eqnarray*}
G(x,t)&=&\operatorname{Im}[\Psi_{1}(x,t)\Psi_{2}^{*}(x,t)]\\
&=&|\Psi_{1}(x,t)||\Psi_{2}(x,t)|\sin[\theta_{1}(x,t)-\theta_{2}(x,t)].
\end{eqnarray*}
The total integrated interband current should be zero, $\int_0^L G(x,t) dx=0$, since there is no net current applied between bands \cite{gurevich2006}. When the two bands are phase-locked, {\it i.e.} $\theta_2(x,t)=\theta_1(x,t)$, $G(x,t)=0$ everywhere. On the other hand, if there are phase textures,  $\int_0^L G^2(x,t) dx\not=0$. Therefore, as a measure of the interband texture, we calculate the time averaged  
$\langle G^2\rangle=\langle\frac{1}{L}\int_0^L G^2(x,t) dx\rangle$ [shown in Fig.\ref{fig:iv}(c)]. 
We find that below $j_c$ the two bands are phase-locked (there are no phase textures in this case), but just above $j_c$ the interband``current"  is finite, $\langle G^2\rangle>0$, indicating that there are phase differences $\theta_2-\theta_1$ induced by the driving current.  The hysteresis in the voltage and in the order parameters is also reflected as a hysteresis in the $\langle G^2\rangle$ vs $J$ curve.


We also plot in  Fig.\ref{fig:psi}(d) the space-time distribution of the interband``current" $G(x,t)$ when there are PSs. Away from the PS the system is phase-locked and thus $G\approx0$. At the time interval between the zeroes of $|\Psi_2(x,t)|^2$ and $|\Psi_1(x,t)|^2$, there is a phase texture where $G\not=0$. 
We note that this phase structure corresponds to a vortex string, which is coreless (there is no vanishing of the order parameter inside) and that it extends along the time direction, so we name it a``time-vortex".
As we will discuss in the upcoming sections, its characteristic time scale,  $t_\gamma$, depends on the driving current and the interband coupling stregth $g$.

\begin{figure}[!t]
	\includegraphics[width=6.5cm]{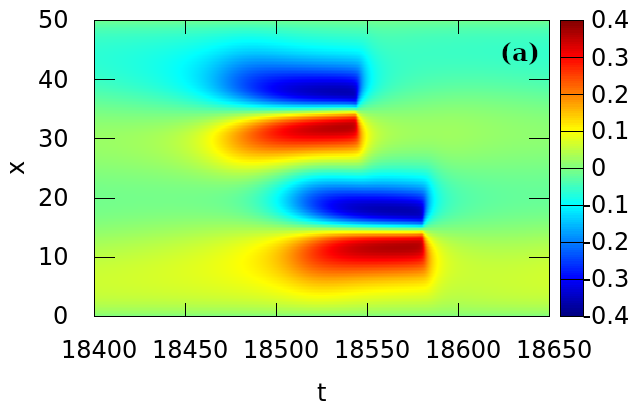}
	\includegraphics[width=6.5cm]{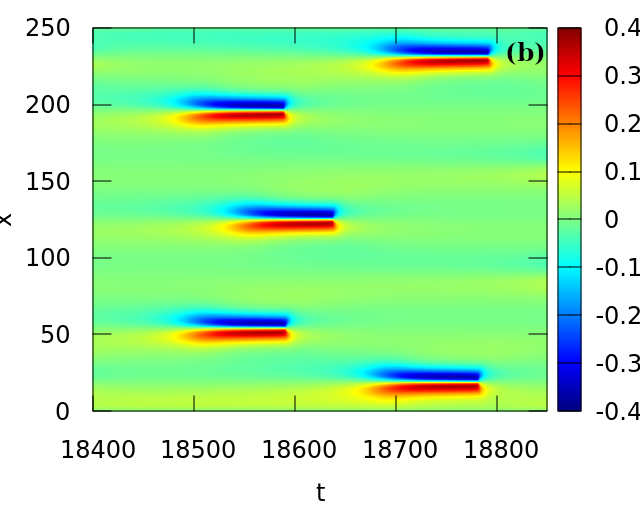}
	\caption{Intensity plots in space-time coordinates of the interband phase texture $G(x,t)=|\Psi_1||\Psi_2|\sin(\theta_1-\theta_2)$, normalized by $\Psi_{10}^2$. Coordinate $x$ is normalized by $\xi_1$ and time by $t_0$. (a) $J=0.3772$ [indicated as $j_B$ in Fig.\ref{fig:iv}], $L=50\xi_1$. (b) $J=0.3731$, $L=250\xi_1$. }
	\label{fig:psis}
\end{figure}

At larger currents (but for $J<j_{2d}$), more PS centers can nucleate. In Fig.\ref{fig:psis}(a) we show the case for a current in the range of the second quasilinear region in the current-voltage curve. We plot $G(x,t)$ showing that there are two centers with time-vortices that connect pairs of PS in the superconducting bands.
Also if the wire length is increased, more slip centers can nucleate in the wire, as seen in  Fig.\ref{fig:psis}(b) for a wire of length $L=250\xi_1$.

\section{Topological  connection in time of the phase slips}

\begin{figure}[!htb]
	\includegraphics[width=\columnwidth]{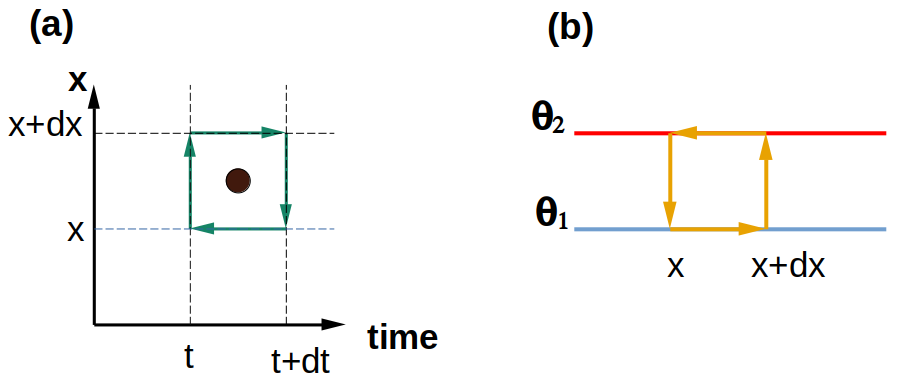}
	\includegraphics[width=\columnwidth]{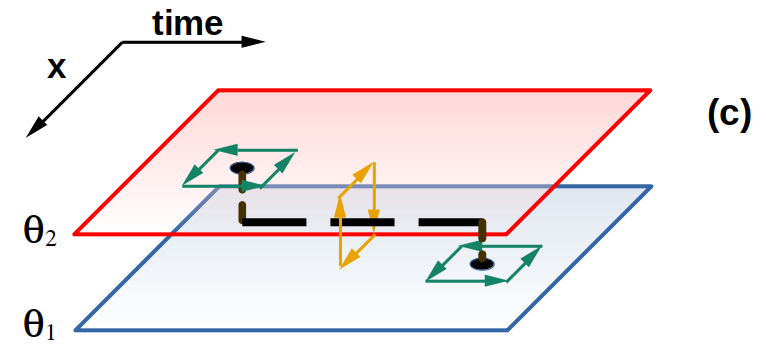}
	\caption{(a) Space-time loop for the calculation of the phase-slip vorticities $\nu_i(x,t)$, as defined in Eq.(\ref{eq:n_i}). (b) Space-interband loop for the calculation of the time-oriented interband vorticity $\nu_\gamma(x,t)$, as defined in Eq.(\ref{eq:n_g}).  (c) Schematic representation of the topological structure of a time-connected phase slip. The dimension perpendicular to the plane $xt$ is an abstract coordinate that assumes the values $\theta_1(x,t)$ in the lower plane and $\theta_2(x,t)$ in the upper plane.}
	\label{fig:schematic}
\end{figure}

Following Ivlev and Kopin \cite{ivlev1978}, we can regard a PS  as a 
vortex (a topological singularity) in two-dimensional space-time. In each $i$-band, the sum of phase differences $\delta_x\theta_i$ and $\delta_t\theta_i$ along a closed loop in space-time must equal an integer multiple of $2\pi$. Accordingly, we define the PS vorticity as 
\begin{equation}
\oint \frac{\partial \theta_i}{\partial {\vec \rho}}\cdot d{\vec \rho}={2\pi}\nu_i
\label{eq:vorps}
\end{equation}
with the space-time coordinate ${\vec \rho}=(x,t)$.

We can calculate  local vorticities $\nu_i(x,t)$ considering infinitesimal loops along closed  paths corresponding to the discretization distances $dx,dt$. In this case, the closed space-time line integral is obtained as a oriented sum along the four segments of the loop  shown in Fig.\ref{fig:schematic}(a).
In the spatial segments we obtain the phase differences $\delta_x\theta_i(x,t)$ from $x+dx$ to $x$ at $t$ fixed;  and in the temporal segments we obtain the phase differences $\delta_t\theta_i(x,t)$ from $t+dt$ to $t$ at $x$ fixed. To calculate numerically the vorticity \cite{teitel1993,phillips2015}, the phase differences $\delta_\mu\theta_i$ ($\mu=x,t$) are  redefined in the $(-\pi,\pi)$ interval as
$\delta_\mu\theta_i \rightarrow [\delta_\mu\theta_i]=\delta_\mu\theta_i+2\pi m_{i,\mu}$, where the link integers are $m_{i,\mu}=-{\rm nint}(\frac{\delta_\mu\theta_i}{2\pi})$, with ${\rm nint}(u)$ the nearest integer of $u$.  We then obtain,

\begin{eqnarray}
	\nu_{i}(\tilde{x},\tilde{t}) & = &
	m_{i,x}(x,t)+m_{i,t}(x+dx,t)\nonumber\\
&  &-m_{i,x}(x,t+dt)-m_{i,t}(x,t) \label{eq:n_i} 
\end{eqnarray}
where the point ($\tilde{x},\tilde{t}$) is located in the interior of the considered rectangle. The integers $m_{i,x}(x,t)$ are defined in the {\it directed} link
between $(x,t)$ and $(x+dx,t)$, meaning that when the summation path  goes from  $(x+dx,t)$ to $(x,t)$ they are added as $-m_{i,x}(x,t)$; similar convention is taken for the $m_{i,t}(x,t)$ integers.

In the two-band superconductor one can also define a vortex string (i.e.  a ``Josephson" vortex \cite{dremov2019}), which is a $2\pi$-singularity in the phase difference $\Delta\theta=\theta_2-\theta_1$.  In this way, it is possible to quantify \cite{comment} an interband vorticity taking a closed loop that goes from a point $x_a$ to $x_b$ in the 1-band and returns from $x_b$ to $x_a$ in the 2-band,  as
\begin{equation}
	2\pi \nu_{\gamma}=\int_{x_a}^{x_b}\delta_x\theta_{1}+\Delta\theta(x_b)-\int_{x_a}^{x_b}\delta_x\theta_{2}-\Delta\theta(x_a)\;.\label{eq:n_g1}
\end{equation}

We can then calculate a local vorticity in a closed small path that goes from $x$ to $x+dx$ and back, see Fig.\ref{fig:schematic}(b), as
\begin{eqnarray}
	\nu_{\gamma}(\tilde{x},t) & = & m_{1,x}(x,t)+m_\Delta(x+dx,t)
	\label{eq:n_g}\\
	&  &- m_{2,x}(x,t)-m_\Delta(x,t)\nonumber
\end{eqnarray}
where we have also restricted to the $(-\pi,\pi)$ interval the interband phase differences  $\Delta\theta$, redefined as
$\Delta\theta \rightarrow [\Delta\theta]=\Delta\theta+2\pi m_{\Delta}$, with the link integers  $m_{\Delta}=-{\rm nint}(\frac{\Delta\theta}{2\pi})$.

\begin{figure}[!htb]
	\includegraphics[width=\columnwidth]{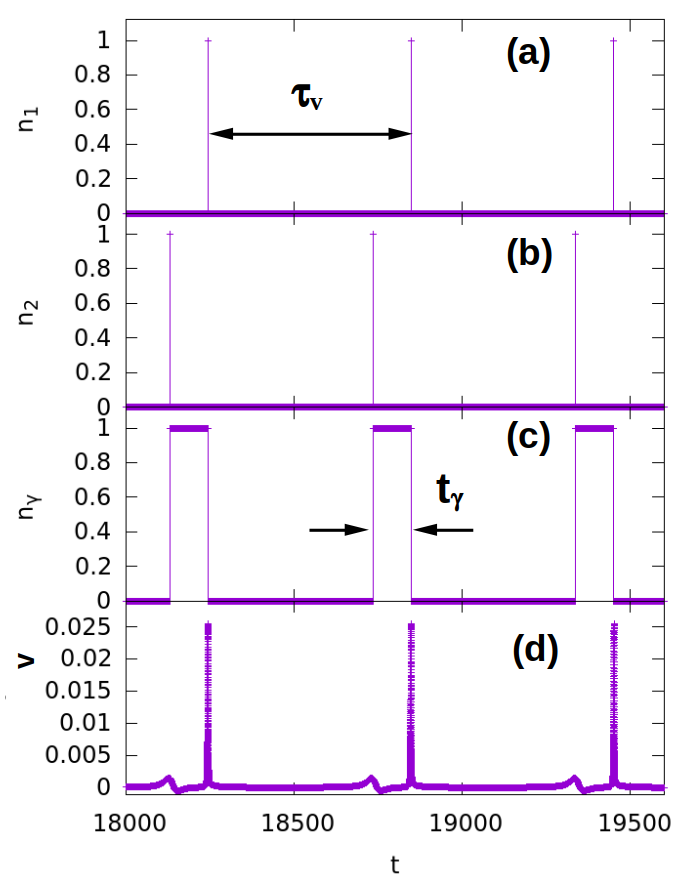}
	\caption{Time-dependence of the t-PS for a current $J=0.372$ [indicated as $j_A$ in Fig.\ref{fig:iv}(a)] for a wire with $L=50\xi_1$. Phase-slip vorticities (a) $n_1$ and (b) $n_2$. (c) Interband vorticity $n_\gamma$. (d)  Voltage per unit length $v$, normalized by $v_0/\xi_1$. Time is normalized by $t_0$.}
	\label{fig:vorticities}
\end{figure}

We also define the total vorticities along the wire at a given time, both the phase-slip vorticities $n_{i}(t)=\sum_{x} \nu_{i}(x,t)$ and the interband vorticities $n_{\gamma}(t)=\sum_{x} \nu_{\gamma}(x,t)$. From Eq.\ (\ref{eq:n_i}), we can write
\begin{eqnarray}
	n_{i}(\tilde{t}) & = &\sum_x\left\{ m_{i,x}(x,t)-m_{i,x}(x,t+dt)\right\} \label{eq:nsum}\\
&+&m_{i,t}(L,t)- m_{i,t}(0,t)
 \nonumber
\end{eqnarray}
where the sums are taken over the $L/dx$ segments in the grid. The second line in Eq.\ (\ref{eq:nsum}) has the same value for the two bands, due to the boundary conditions Eq.(\ref{BC}) at $x=0$ and $x=L$.
Similarly, from Eq.(\ref{eq:n_g}) we obtain,
\begin{eqnarray}
	 n_\gamma (\tilde{t}) 
	& = &\sum_x \left\{m_{2,x}(x,t)-m_{1,x}(x,t)\right\} \label{eq:ngsum}\\
	&+& m_\Delta(L,t)-m_\Delta(0,t)\,, \nonumber
\end{eqnarray}
where the second line vanishes since $\theta_2=\theta_1$ at $x=0$ and $x=L$
due to Eq.(\ref{BC}).
 Combining Eqs.(\ref{eq:nsum}) and (\ref{eq:ngsum}),
  it follows that
 \begin{equation}
 n_\gamma (\tilde{t}+dt)-n_\gamma (\tilde{t})=n_2(\tilde{t})-n_1(\tilde{t}).
 \label{eq:cont}
 \end{equation} 

We plot in Fig.\ref{fig:vorticities}(a),(b),(c) the total vorticities as functions of the time. We see that, periodically, 
 a phase-slip is created first in the``weak band" when $n_2=1$ at an instant of time 
 (while $n_1=n_\gamma=0$).
Right after this  time, the interband vorticity becomes $n_\gamma=1$,  consistent with Eq.(\ref{eq:cont}). Later on, at the end of a finite period of time $t_\gamma$ (during which $n_1=n_2=0$), a phase-slip is created in the ``strong band" when $n_1=1$, and $n_\gamma$ switches back to zero. Therefore, there is a topological continuity between the two PS, which are connected along the time line of length $t_\gamma$.
  In other words, the two phase-slips are connected across time by an interband vortex of length $t_\gamma$. To emphasize this characteristic, we will call this structure``time-connected phase slips" (t-PS). In Fig.\ref{fig:schematic}(c) we show a schematic representation of a t-PS. This process  repeats periodically, with a period $\tau_V \propto 1/V > t_\gamma$. We can see  three instances of t-PS in the time window shown in Fig.\ref{fig:vorticities}.

 The t-PS are characterized by a two-peak structure in the time dependence of the voltage $v(t)$, as can be seen in Fig.\ref{fig:vorticities}(d). There is first a shallow peak that corresponds to the PS in the weak band, and after the time $t_\gamma$ there is a sharp peak that corresponds to the PS in the strong band. This agrees with the electric field dependence shown in Fig.\ref{fig:psi}(c).

\begin{figure}[!htb]
	\includegraphics[width=\columnwidth]{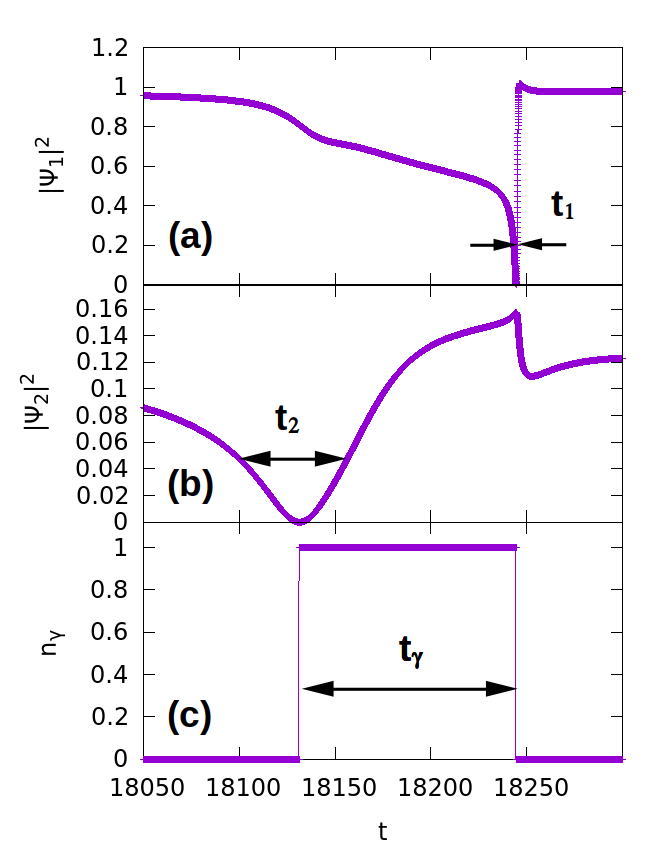}
	\caption{Structure of a t-PS. Time dependence of (a) $|\Psi_1|^2$, (b) $|\Psi_2|^2$ and (c) $\nu_\gamma$, at $x=L/2=25\xi_1$ for $J=0.372$ . For this current density, $\nu_\gamma (L/2,t)=n_\gamma (t)$. Time is normalized by $t_0$ and order parameters by  $\Psi_{10}^2$. }
	\label{fig:psi12}
\end{figure}

In Fig.\ref{fig:psi12} we further analyze  the structure of the time scales that characterize a t-PS. We plot $|\Psi_{1}|^2$, $|\Psi_{2}|^2$ and $n_{\gamma}$ as functions
of time for $J=0.372$, at the point of nucleation, $x=L/2$. We see that 
$|\Psi_{2}|^2$ vanishes first and  $|\Psi_{1}|^2$ vanishes at a later time  . The time scale for the zero of $|\Psi_{1}|^2$ is $t_1$ and the time scale for $|\Psi_{2}|^2$ is $t_2 > t_1$. In the interval between the two zeroes the interband vorticity is $n_{\gamma}=1$ (and $n_{\gamma}=0$ outside this interval) during a time $t_\gamma$. From the analysis of Fig.\ref{fig:vorticities} and Fig.\ref{fig:psi12} we infer that the inequalities $\tau_V> t_\gamma > t_2 > t_1$ have to be fulfilled in order to have a well-defined periodic sequence of t-PS.

\begin{figure}[!htb]
	\includegraphics[width=8cm]{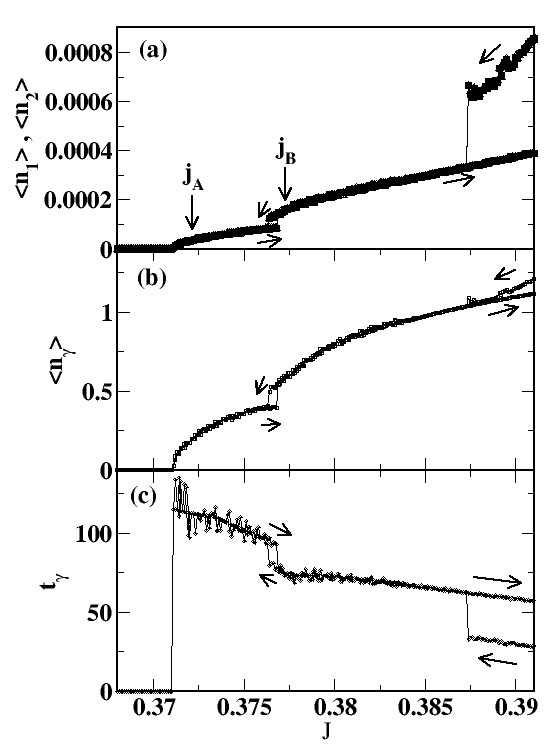} \caption{(a) Average number of phase slips in each of the two bands. Square symbols represent $\langle n_{1}\rangle$ and triangles represent $\langle n_{2}\rangle$. They are difficult to distinguish in the plot since both values coincide. (b) Average number of interband time-vortices, $\langle n_{\gamma}\rangle$. (c) Estimated average time length, ${\bar t_\gamma}=\delta t\langle n_{\gamma}\rangle/\langle n_{i}\rangle$, normalized by $t_0$. Current $J$ is normalized by $j_0$.}
	\label{fig:nvsj}
\end{figure}

In Fig.\ref{fig:nvsj} we study,
as a function of $J$, the average number of phase slips per step, $\langle n_{i}\rangle=\frac{1}{N_T}\sum_{x,t}\nu_{i}(x,t) $ (with $N_T$ the number of time steps, and the time-averaged total number of interband vortices $\langle n_{\gamma}\rangle=\frac{1}{N_T}\sum_{x,t}\nu_{\gamma}(x,t) $.
We find that, as expected, the number of PS is the same in each band, and that they agree with the average voltage, $\langle n_{1}\rangle=\langle n_{2}\rangle \sim V/2\pi$.
We see in Fig.\ref{fig:nvsj} (b) that  simultaneously with the nucleation of PS above $j_c$, there is a finite  interband vorticity $\langle n_{\gamma}\rangle$ that increases with the current.

More interestingly, we can quantify the average time length of the t-PS. We estimate numerically the average time-length of time-vortices as
${\bar t_\gamma}=\delta t\langle n_{\gamma}\rangle/\langle n_{i}\rangle$.
($\delta t\langle n_{t}\rangle $  is total time-length of time-vortices, and ${\bar t_\gamma}$ is the time-length per t-PS).
When there is only one PS center in the sample, with vorticity $1$, ${\bar t_\gamma}$ is the time-length of the t-PS. We plot in Fig.\ref{fig:nvsj} (c) the dependence of ${\bar t_\gamma}$ on the current. When approaching the critical current $j_c$ from above the length ${\bar t_\gamma}$ increases, reaching its largest value at $J\rightarrow j_c^+$. For  $J<j_c$ there are no time-vortices and their length becomes meaningless;  we adopt the convention to define ${\bar t_\gamma}=0$ whenever $\langle n_{\gamma}\rangle=0$,  to emphasize the difference with the case of  phase solitons \cite{tanaka2001,gurevich2003}, for which $\langle n_{\gamma}\rangle\not=0$ and $\langle n_{i}\rangle=0$, implying that ${\bar t_\gamma}$ is infinite.

\section{Dependence on interband coupling and relaxation parameters}


To study the dependence of the t-PS on the interband coupling $g$, it is more convenient to work at a fixed voltage $V$ than at fixed current $J$. The current $j_c$  for the onset of voltage changes when changing the system parameters. If $J$ is fixed and  $g$ is varied, the voltage will change with $g$ due to the dependence of $j_c$ with $g$. On the other hand, by fixing the voltage $V$, we are guaranteed to work at a fixed time-scale, given by the expected time period for phase slips, $\tau_V\propto 1/V$. 

Numerically, we impose the fixed voltage $V$  by solving the second order equation for $\phi(x,t)$:
\begin{equation}
	-\frac{\partial^2\phi}{\partial x^2} +\frac{\partial}{\partial x}\left[\operatorname{Im}\left(\Psi_{1}^{*} \nabla \Psi_{1}\right)+k \operatorname{Im}\left(\Psi_{2}^{*} \nabla \Psi_{2}\right)\right]=0\;,
\end{equation}
with the boundary condition $\phi(0,t)=0$,  $\phi(L,t)=-VL$.
In this case the current  depends on time, and  it can be calculated as 
$J(t)=V +\frac{1}{L}\int_0^L[\operatorname{Im}\left(\Psi_{1}^{*} \nabla \Psi_{1}\right)+k \operatorname{Im}\left(\Psi_{2}^{*} \nabla \Psi_{2}\right)]dx$.
Simultaneously, the TDGL equations for $\Psi_i(x,t)$  are solved with the superconducting bank boundary conditions, as in the previous sections.

\begin{figure}[!htb]
	\includegraphics[width=8cm]{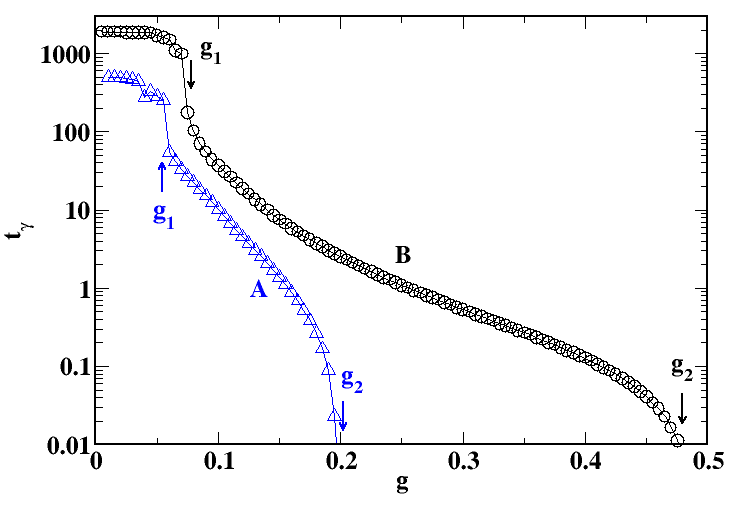}
	\caption{Estimated average time length ${\bar t_\gamma}$ (normalized by $t_0$) as a function of the interband coupling parameter $g=\gamma/|\alpha_1|$. Parameter sets A: $(V=0.00745, a=0.2, b=0.7, k=5.2, d=10, \eta_1=0.07)$,
		and B: $(V=0.002, a=0.2, b=1.2, k=5.2, d=2, \eta_1=0.5)$. The limits $g_1$ and $g_2$ for the existence of t-PS are indicated.}
	\label{fig:tcps}
\end{figure} 

 We fix a low voltage, near the onset of dissipation, and plot in Fig.\ref{fig:tcps} the average time-length of the t-PS, ${\bar t_\gamma}$, as a function of the interband coupling $g$,
for two different choices of parameters, both in the case $t_2/t_0=\eta_2/a>1$ and $t_1/t_0=\eta_1< 1$.
Ref.\cite{gurevich2003} estimated the time scale for variations in the interband phase as $t_{\rm interband}=\eta_1\eta_2\Delta_1\Delta_2/[g(\Delta_1^2\eta_1+\Delta_2^2\eta_2)]$ for the case of low current, $J\ll j_c$ (no dissipation, $v=0$), in the limit of small coupling, $g\ll a$; $\Delta_1$ and $\Delta_2$ are the equilibrium order parameters. Fig.\ref{fig:tcps}  shows that ${\bar t_\gamma}$  decreases with increasing $g$, as expected from the estimate of $t_{\rm interband}$, but the dependence is more complex since, among other factors, $\Delta_1$ and $\Delta_2$ depend on $g$, and  on the applied voltage.

We find that the t-PS can occur in an interval $g_1<g<g_2$. Below a coupling $g_1$ the value of ${\bar t_\gamma}$ saturates to a constant,  
while at $g_2$ the time 
${\bar t_\gamma}$ vanishes. The lower limit $g_1$ corresponds to the condition that $t_\gamma $ has to be smaller than the period of the PS (the t-PS do not overlap), $t_\gamma < \tau_V $. The value of $g_1$ is reached when  $t_\gamma \approx \tau_V \approx 2\pi/V  $. Indeed, we find that 
the saturation value of ${\bar t_\gamma}$ grows as $1/V$.
   
The upper limit $g_2$ stems from requirement that 
 $t_\gamma$  has to be larger than the smallest relaxation
time of 
the order parameters, in this case $t_\gamma >  t_{1}=\eta_1$. If this  condition is not fulfilled, the PS of the two order parameters are effectively synchronized. 
 Fenchenko and Yerin \cite{fenchenko2012} studied PS in two-band superconducting wires, finding synchronized PS. Indeed, the TDGL parameters studied in their work correspond to the $g>g_2$ case of our analysis.

We can also see in Fig.\ref{fig:tcps} that for larger relaxation parameter $\eta_1$ (always under the condition $t_2>t_1$) the time length $t_\gamma$ is larger, as expected from the above mentioned expression for $t_{\rm interband}$. For $\eta_1>1$, however, the upper limit $g_2$ decreases significantly, and the region $g_1<g<g_2$ where the t-PS could be generated  becomes too small.

\section{Discussion}

We have found a new topological object in driven one-dimensional two-band superconductors. Above the critical current we find that there are separate nucleations of PS in each band that are displaced in time and topologically connected through a vortex string oriented along the time direction.  It is interesting to note that the t-PS are analogous to kinked vortices in layered superconductors \cite{feinberg1990,ivlev1990,blatter1994} (see for instance the Fig.35 of \cite{blatter1994}). In kinked vortices there are two-dimensional space coordinates instead of space-time coordinates, and the 2D pancake vortices are the analogues of the PS. However, even though the topological structures of t-PS and kinked vortices are similar, their physics and dynamical behavior is, of course, very different.

In single-band one dimensional superconductors, the possible solutions depend on one free parameter of the TDGL equation (the relaxation rate $\eta$), plus the length of the wire $L$, the boundary conditions, and the driving current $J$ \cite{michotte2004,vodolazov2011,ludac2008,kim2010,baranov2011,baranov2013,kallush2014,berger2015,kimmel2017}.
In the case of two-band superconductors, there are {\it six} free parameters in the TDGL equations ($a,b,k,g,\eta_1,\eta_2$), in addition to wire length,  boundary conditions and driving current $J$. Therefore, there is a large variety of behaviors one could find in this problem. Besides the parameter sets shown here, we have explored several other choices of the TDGL equation parameters \cite{comment2} within the ranges
$0<a < 1$, $b>a$, $k>a$, $g<a$, $\eta_2/a>\eta_1$ .
(Note that after exchange  of the band index $1 \rightarrow 2$ in the TDGL equations, the parameters change as $a \rightarrow 1/a$ , $b \rightarrow 1/b$, $k  \rightarrow 1/k$, $ g \rightarrow g/a$ $\eta_1 \rightarrow \eta_2ak/b$, $\eta_2 \rightarrow \eta_1ak/b$, and the same results are obtained.)
We have also explored two other  possible boundary conditions: (i) $\partial \Psi_i/\partial x=0$ at $x=0,L$, and (ii) $\Psi_i=0$ at $x=0,L$; and system sizes in the range $L/\xi_1=10-250$.
 
We have found that the nucleation of t-PS occurs for most of the parameters  explored,  always within some range $g_1<g<g_2$, and in a narrow range of currents near a critical current $j_c$. The range in $g$ depends on the parameters $a,b,k,\eta_1,\eta_2$ and the voltage $V$, as shown in Fig.\ref{fig:tcps} for two particular cases. The occurrence of  t-PS is independent of the boundary conditions, i.e.\ it is a bulk phenomenon.  Different boundary conditions give slightly different history dependencies in the current-voltage curve, but in all cases there are t-PS above an onset current $j_c$.

When decreasing the current, approaching the critical current from above, the time length $t_\gamma$ increases. One could think that when $J\rightarrow j_c^+$, $t_\gamma\rightarrow\infty$. Even though we do not discard this scenario, we do not find such a divergence for the range of currents explored. Another related question is if there is a relationship between the existence of the t-PS for  $J>j_c$ and the existence of phase solitons for $J<j_c$. In principle, a t-PS with $t_\gamma=\infty$ is a phase soliton, {\it i.e.},  there is a phase texture with $\theta_1\neq\theta_2$ at a given point in space for all times \cite{gurevich2003,gurevich2006,tanaka2001}. Or one could interpret the periodic sequence of t-PS as a `dismembered' phase soliton which is interrupted by PS and broken in several segments along the time direction.
We have found that for some sets of  TDGL parameters it is possible to have both, phase solitons for $J<j_c$ and t-PS for $J>j_c$, but this is not always the case. In general, there is no direct connection between the conditions for the existence of these two topological objects.
For the two sets of TDGL parameters shown in this paper there are no phase solitons for $J<j_c$, and there are t-PS above $j_c$.

For increasing current, there are jumps to different dynamical regimes in the current-voltage curve, as seen for instance in Fig.\ref{fig:iv}. At first, we find regimes with one, two, three t-PS centers. The maximum number of t-PS centers obtained depends on the length $L$. At larger currents other dynamical regimes are found: quasiperiodic regimes with two or more t-PS centers each with different periodicity, chaotic regimes  (similar to what was reported in \cite{fenchenko2012}),  a regime where in one part of the wire the 2-band is normal, and on top of it there are PS of the 1-band, with the size of the 2-band normal sector growing with current, etc.   Which of these high-current regimes are observed, depends on the choices of the TDGL parameter sets. Here we have focused on describing the t-PS and their topological structure, and leave for future work the  study of the very rich variety of dynamical behaviors that can be found in current driven two-band superconducting wires.

A possible experimental evidence of t-PS would be the measurement of a two-peak structure  in the time dependence of the voltage near the critical current. This requires a resolution in the time scale of $t_\gamma \sim 10-100 t_0$. Since $t_0\sim 10^{-11}-10^{-10} s$ in conventional superconductors \cite{kadin1986}, this can be very challenging to achieve.
Besides two-band superconductors like MgB$_2$ and iron compounds, the t-PS  can also be observed in   artificially fabricated structures of Josephson-coupled bilayer superconductors \cite{bluhm2006,ishizu2023}, where  parameters could be tuned by using  layers with different mean free path, or layers with different superconductors, or varying the layer and interlayer thicknesses. 

\section*{Acknowlegments}
DD acknowledges support from CNEA, CONICET , ANPCyT ( PICT2019-0654) and  UNCuyo (06/C026-T1).

\appendix

\section{Crank-Nicholson  algorithm for the two-band TDGL equations}

To solve the TDGL equations we adapt to the one-dimensional two-component superconductor the semi-implicit Cranck-Nicholson algorithm described in \cite{winiecki2002}.

In the $x$ coordinate we use the discretization,
$$
\frac{\partial^2\Psi(x, t)}{\partial x^2}  \rightarrow \frac{\Psi(x+d x, t)+\Psi(x-d x, t)-2 \Psi(x, t)}{d x^{2}}
$$

For a one-component superconductor, the Crank-Nicholson method discretizes the time dependence of the equation
$$
\left(\frac{\partial}{\partial t}+i \phi\right) \Psi(x, t)=D(x, t)
$$
as:
$$
\frac{\Psi(x, t+d t)-e^{-i \phi d t} \Psi(x, t)}{d t}=\frac{1}{2}[D(x, t)+D(x, t+d t)]
$$
with 
$$D(x,t)=\frac{1}{\eta}\left(\frac{\partial^2\Psi(x, t)}{\partial x^2}+(1-|\Psi(x, t)|^{2}\right)$$
 and $d t$ is the time step discretization.

In  the nonlinear term of $D(x, t+d t)$ we approximate $|\Psi(x,t+dt)|^{2}\approx|\Psi(x, t)|^{2}$. In this way, it is possible to rewrite the equations to obtain $\Psi(x,t+dt)$  as  a function $\Psi(x,t)$ solving the  tridiagonal form, 
\begin{eqnarray}
&&A_{n} \Psi_{n}(t+d t)+B \Psi_{n+1}(t+d t)+B \Psi_{n-1}(t+d t)\nonumber\\
&&=f_n(t)
\end{eqnarray}
with the coefficients $A_{n}=1+\frac{dt}{2\eta}(\frac{2}{dx^2}-1+|\Psi_{n}(t)|^2)$ and $B=-\frac{dt}{2\eta dx^2}$. The right hand side term is $f_n(t)=\Psi_{n}(t)e^{-i\phi_n(t)dt}+\frac{dt}{2}D_n(t)$. The subindex $n$ indicates the discretized spatial coordinate, $ \Psi_{n}(t)=\Psi(n d x, t)$. For a systen of length $L$, we define $N$ as  $d x N=L$.
We solve the tridiagonal equation with the recurrence:
$$
\Psi_{n-1}(t+d t)=\alpha_{n} \Psi_{n}(t+d t)+\beta_{n}
$$
where
\begin{eqnarray}
\alpha_{n+1}&=&-\frac{B}{B \alpha_{n}+A_{n}}\\
 \beta_{n+1}&=&\alpha_{n+1}(\beta_{n}-\frac{f_{n}}{B})
\end{eqnarray}
In the case of Dirichlet boundary conditions that fix $\Psi(x=0)=\Psi_0$ and $\Psi(x=L)=\Psi_N$, the recurrence equations for $\alpha_n$ and $\beta_n$ are started with
$\alpha_1=0$, $\beta_1=\Psi_0$. Once obtained  $\alpha_n$ and $\beta_n$, the recurrence equation for $\Psi_{n}$ is started from  $n=N$ with the boundary value $\Psi_N$.
 
In the two-band case, the Crank-Nicholson method  leads to two coupled tridiagonal equations \cite{casali2013},
$$A_{1,n} \Psi_{1,n}+B_1\Psi_{1,n+1}+B_1 \Psi_{1,n-1}+C_1\Psi_{2,n}=f_{1,n}(t)$$
$$A_{2,n} \Psi_{2,n}+B_2\Psi_{2,n+1}+B_2 \Psi_{2,n-1}+C_2\Psi_{1,n}=f_{2,n}(t)$$
with $\Psi_{1,n}$, $\Psi_{2,n}$ in the left hand side evaluated at $t+dt$. The coefficients are
$A_{1,n}=1+\frac{dt}{2\eta_1}\left(\frac{2}{dx^2}-1+|\Psi_{1,n}(t)|^2\right)$,
$A_{2,n}=1+\frac{dt}{2\eta_2}\left(\frac{2k}{dx^2}-a+b|\Psi_{2,n}(t)|^2\right)$,
 $B_1=-\frac{dt}{2\eta_1 dx^2}$,
  $B_2=-\frac{k dt}{2\eta_2 dx^2}$,
  $C_1=-\frac{dt}{2\eta_1}g$, 
  $C_2=-\frac{dt}{2\eta_2}g$. The right-hand side terms are
   $f_{i,n}(t)=\Psi_{i,n}(t)e^{-i\phi_n(t)dt}+\frac{dt}{2}D_{i,n}(t)$; with $D_1(x,t)=\frac{1}{\eta_1}\left(\frac{\partial^2\Psi_1(x, t)}{\partial x^2}+(1-|\Psi_1(x, t)|^{2}\right)$, and
$D_2(x,t)=\frac{1}{\eta_2}\left(k\frac{\partial^2\Psi_2(x, t)}{\partial x^2}+(a-b|\Psi_2(x, t)|^{2}\right)$.

The recurrences are
$$\Psi_{1,n-1}=\alpha_{1,n} \Psi_{1,n}+\beta_{1,n}+\delta_{1,n} \Psi_{2,n}$$
$$\Psi_{2,n-1}=\alpha_{2,n} \Psi_{2,n}+\beta_{2,n}+\delta_{2,n} \Psi_{1,n}$$
with
$$
\begin{aligned}
&\alpha_{1,n+1}=-\frac{B_1}{h_n}(A_{2,n}+B_2\alpha_{2,n})\\
&\alpha_{2,n+1}=-\frac{B_2}{h_n}(A_{1,n}+B_1\alpha_{1,n})\\
&\delta_{1,n+1}=\frac{B_2}{h_n}(C_1+B_1\delta_{1,n})\\
&\delta_{2,n+1}=\frac{B_1}{h_n}(C_2+B_2\delta_{2,n})
\end{aligned}
$$
$$
\begin{aligned}
&\beta_{1,n+1}=\alpha_{1,n+1}(\beta_{1,n}-\frac{f_{1,n}}{B_1})+\delta_{1,n+1}(\beta_{2,n}-\frac{f_{2,n}}{B_2})\\
&\beta_{2,n+1}=\alpha_{2,n+1}(\beta_{2,n}-\frac{f_{2,n}}{B_2})+\delta_{2,n+1}(\beta_{1,n}-\frac{f_{1,n}}{B_1})\;,\\
\end{aligned}
$$
with $h_n=(A_{1,n}+B_1\alpha_{1,n})(A_{2,n}+B_2\alpha_{2,n})-(C_1+B_1\delta_{1,n})(C_2+B_2\delta_{2,n})$.
For the boundary conditions that fix $\Psi_i(x=0)=\Psi_{i,0}$ and $\Psi_i(x=L)=\Psi_{i,N}$, the recurrence equations for the $\alpha$'s, $\beta$'s and $\delta$'s are started with
$\alpha_{1,1}=\alpha_{2,1}=\delta_{1,1}=\delta_{2,1}=0$, $\beta_{i,1}=\Psi_{i,0}$. Once obtained the  $\alpha$'s, $\beta$'s and $\delta$'s, the recurrence equations for $\Psi_{i,n}$ are started from  $n=N$ with the boundary values $\Psi_{i,N}$.

Standard Euler and Runge-Kutta algorithms require small  $d t$ integration steps for  stability, since the TDGL equations are of diffusion type. In our case, we need a $dt \sim 0.001$ to achieve stability in an Euler  algorithm.
The advantage of the Crank-Nicholson algorithm is its stability for large time steps \cite{winiecki2002}. In our case we have verified that for  up to $dt = 0.5$ we can have numerically stable solutions. Here we solve the TDGL equations with $dt=0.02$ for better accuracy.
In our simulations we have integrated the equations allowing for equilibration for each value of $J$ (or $V$ in the fixed voltage case) in the first $10^6$ time steps, and time averages are computed for the following $10^5$ steps.

\section{Gauge-invariant formulation of PS vorticity}

The Eq.(\ref{eq:vorps}) has to be rewritten in a gauge invariant form, to properly take into account electric potential variations \cite{ivlev1978}. Defining the gauge invariant space-time vectors $${\vec q}_i = (\frac{\partial\theta_i}{\partial x}-\frac{2e}{\hbar c}A_x,\frac{\partial\theta_i}{\partial t}+\frac{2e}{\hbar}\phi),$$ we obtain the Ivlev and Kopnin result \cite{ivlev1978} in its complete form,  and generalized for multiband superconductors, as
$$
\oint {\vec q}_i\cdot d{\vec \rho}={2\pi}\nu_i-\frac{2e}{\hbar}\int_S E\cdot d\sigma,
$$
where $E$ is the electric field and the integral $\int_S$ is in the space-time surface enclosed by the loop, with  $d\sigma=dxdt$. This is valid for any space-time loop. For a large loop that extends of for a long total time $T$ and along the whole length  $L$ of the wire we can consider that the left-hand side vanishes, and we obtain the quantization condition \cite{ivlev1978}
$${2\pi}N_i =\frac{2e}{\hbar} \langle E\rangle LT\,, $$
where $N_i$ is the total number of PS in the $i$-band in the time interval $T$, and $\langle E\rangle$ is the space-time averaged electric field. Therefore, the average electric field is proportional to the average number of PS per unit time, $\langle n_i\rangle = N_i/T$. Since the expression at the right of equality is the same for both bands, we deduce that there must be the same total number of PS in the two bands, ${\it i.e.}$, $N_1=N_2$.

For an infinitesimal loop along the discretization distances $dx,dt$ [as shown in Fig.\ref{fig:schematic}(a)], we define the gauge-invariant space and time discrete phase differences as
\begin{eqnarray}
 \delta_x\theta_i(x,t)&=&\theta_{i}(x+dx,t)-\theta_{i}(x,t)-A_x(x,t) \nonumber\\ \delta_t\theta_i(x,t)&=&\theta_{i}(x,t+dt)-\theta_{i}(x,t)+\phi(x,t)dt. \nonumber
\end{eqnarray}

 Then,  to calculate numerically the local vorticities in the standard way \cite{teitel1993,phillips2015,smorgrav2005}, redefining each of the local gauge invariant differences in the $(-\pi,\pi)$ interval  we obtain:
\begin{eqnarray}
	2\pi \nu_{i}(\tilde{x},\tilde{t}) & = & [\delta_x\theta_i(x,t)]+[\delta_t\theta_i(x+dx,t)]-\label{eq:n_ib}\\
	&  & [\delta_x\theta_{i}(x,t+dt)]-[\delta_t\theta_{i}(x,t)]+\nonumber \\
	&  & \left(\phi(x,t)-\phi(x+dx,t)\right)dt	 \nonumber
\end{eqnarray}
where, for a phase $\alpha$, $[\alpha]$ is taken in the interval $(-\pi,\pi)$, numerically calculated as $[\alpha]=\alpha-2\pi{\rm nint}(\frac{\alpha}{2\pi})$, with ${\rm nint}(u)$ the nearest integer of $u$. The last term corresponds to the local electric field  $\phi(x,t)-\phi(x+dx,t)=E(x,t)dx$, and the point ($\tilde{x},\tilde{t}$) is located in the interior of the considered plaquete. In terms of the link integers, the expression (\ref{eq:n_i}) stays  the same 
\begin{eqnarray}
	\nu_{i}(\tilde{x},\tilde{t}) & = &
	m_{i,x}(x,t)+m_{i,t}(x+dx,t)\nonumber\\
	&  &-m_{i,x}(x,t+dt)-m_{i,t}(x,t) \,.
\end{eqnarray}
The link integers in their gauge invariant form are
\begin{eqnarray}
	 m_{i,x}(x,t)&=&-{\rm nint}\left(\frac{\theta_{i}(x+dx,t)-\theta_{i}(x,t)-A_x(x,t)}{2\pi}\right)\nonumber\\
	  m_{i,t}(x,t)&=&-{\rm nint}\left(\frac{\theta_{i}(x,t+dt)-\theta_{i}(x,t)+\phi(x,t)dt}{2\pi}\right)\nonumber
\end{eqnarray}
In this work, we have calculated numerically the vorticities using the above expressions in the gauge with $A_x=0$.


\begin{thebibliography}{99}
\bibitem{comment} We note that is also possible to define a space-oriented  interband vorticity as 
 $\nu_{s}(x,\tilde{t})  =  m_{1,t}(x,t)+m_\Delta(x,t+dt)-
 m_{2,t}(x,t))-m_\Delta(x,t)$. 
This corresponds to an interband vorticity in the direction perpendicular to the time-vortex line in Fig.\ref{fig:schematic}(b).
We have found that $\nu_{s}(x,\tilde{t})=0$
in the regimes where there are t-PS, meaning that the PS centers in the two bands always occur in the same point in space. At higher currents, when there are chaotic regimes, or when the 2-band becomes partially or totally normal, we find that $\nu_{s}(x,\tilde{t})\not=0$.
\bibitem{comment2} The parameter sets that we have explored were  within the ranges $a\in(0.1,0.7)$, $b\in(0.7,1.4)$, $k\in(1,6)$, $g\in(0.02,0.4)$, $\eta_1\in(0.07,6)$, $d\in (0.02,10)$ and $\eta_2=kd\eta_1$, in different combinations keeping the relations $b>a$, $k>a$, $g<a$, $kd>a$.
\end{thebibliography}

\end{document}